\documentclass{article}

\usepackage{PRIMEarxiv}
\usepackage[utf8]{inputenc} 
\usepackage[T1]{fontenc}    
\usepackage{hyperref}       
\usepackage{url}            
\usepackage{booktabs}       
\usepackage{amsfonts}       
\usepackage{nicefrac}       
\usepackage{microtype}      
\usepackage{lipsum}
\usepackage{fancyhdr}       
\usepackage{graphicx}       
\usepackage{float}
\usepackage{amsmath}
\graphicspath{{media/}}     

\title{Using Machine Learning to predict Characteristics of Microstrip Line and Microstrip Patch Antenna
}
\author{
  Bharath Balaji \\
  Department of Electronics and
Communication Engineering \\
  National Institute of Technology \\
  Trichy\\
  \texttt{bharath.k.balaji@gmail.com} \\
   \And
  Dr. S. Raghavan \\
  Department of Electronics and
Communication Engineering \\
  National Institute of Technology \\
  Trichy\\
  \texttt{raghavan@nitt.edu} \\
}

\begin{document}
\maketitle

\begin{abstract}
This study, conducted in 2017, explores the use of Machine learning algorithms to predict Characteristics of Transmission Lines such as Impedance or resonance frequency using design parameters of Transmission Lines. Using formulas and equations that define the characteristics of Transmission lines, training data was generated. We trained different models for this dataset. The extent of deviation of predicted output from the actual output was measured in terms of maximum error and average error. This helped determine how well an algorithm worked for a particular transmission line. Further, the best-suited algorithm for each transmission line under consideration was found based on the error. 
\footnote{This research was conducted in the year 2017.}
\end{abstract}

\section{Introduction}
Machine learning has significantly advanced the field of microwave research, particularly in the areas of modeling, simulation, and optimization. Notably, the work by Naser-Moghaddasi et al. \cite{naser-moghaddasi2007} demonstrated the application of heuristic artificial neural networks to analyze and synthesize the performance characteristics of rectangular microstrip antennas. Their approach highlighted the potential of neural networks to predict antenna behavior effectively, thereby streamlining the design process. Building upon such foundational studies, this paper extends the use of machine learning to model a broader range of transmission line characteristics. Specifically, we expand the application of machine learning models to include various types of transmission lines such as Microstrip, Slotline, Stripline, Co-Planar Waveguide (CPW), Co-Planar Strip (CPS), and Microstrip Patch Antenna, employing both Linear Regression and advanced Neural Networks. In this publication, we explore the results for the Microstrip line and Microstrip patch Antenna.

This research leverages standardized historical data to train models that can predict the electrical properties of planar transmission lines, based on physical dimension parameters. This method simplifies the practical design process, enabling the prediction of output parameters for all combinations of inputs before fabrication. Such predictive modeling saves considerable time and resources in antenna design and simulation efforts. Furthermore, the trained models can be utilized to predict the characteristics of new transmission lines, eliminating the need for extensive empirical simulation. This study aims not only to validate the effectiveness of machine learning models in replicating known transmission line behaviors but also to enhance their accuracy and generalizability compared to earlier works. Thus, this research not only confirms the utility of machine learning in this domain but also advances its capability to accommodate a wider array of transmission line configurations and complexities.

\section{Transmission Line Models}
\label{sec:headings}

According to a Microstrip design proposed by K.C.Gupta and Ramesh Garg\cite{gupta1979}: For w/h > 1
\[
\epsilon_{eff} = \frac{\epsilon_r + 1}{2} + \frac{\epsilon_r - 1}{2} \left( \frac{1}{\sqrt{1+12\frac{H}{W}}} \right)
\]

\[
Z_0 = \frac{120\pi}{\sqrt{\epsilon_{eff}}} \left( \frac{W}{H} + 1.393 + \frac{2}{3}\ln\left(\frac{W}{H} + 1.444\right) \right) \Omega
\]

A proposed design of Microstrip Patch Antenna by Bablu Kumar Singh\cite{singh2015}:

\[
\epsilon_{\text{eff}} = \frac{\epsilon_r + 1}{2} + \frac{\epsilon_r - 1}{2} \left(1 + 12\frac{h}{W}\right)^{-\frac{1}{2}}
\]

\[
\Delta L = 0.412h \left( \frac{\epsilon_{\text{eff}} + 0.3}{\epsilon_{\text{eff}} - 0.258} \left(\frac{W}{h} + 0.264\right) \right)
\]

\[
f_r = \frac{c}{2\sqrt{\epsilon_{\text{eff}}}(L + 2\Delta L)}
\]

\subsection{Microstrip Lines}
Microstrip is a planar transmission line used to carry
Electro-magnetic waves (EM waves) or microwave frequency
signals. It consists of 3 layers, conducting strip, dielectric
, and Ground plane. It is used to design and fabricate RF
and microwave components such as directional couplers, power
dividers/combiners, filters, antennae, MMIC, etc.

\(\varepsilon_r\) for all the plots using ANN
The dashed line implies Predicted Output
The dotted line implies Actual Output

\subsection{Results for Microstrip}

\begin{figure}[H]
  \centering
  \includegraphics[scale=0.4]{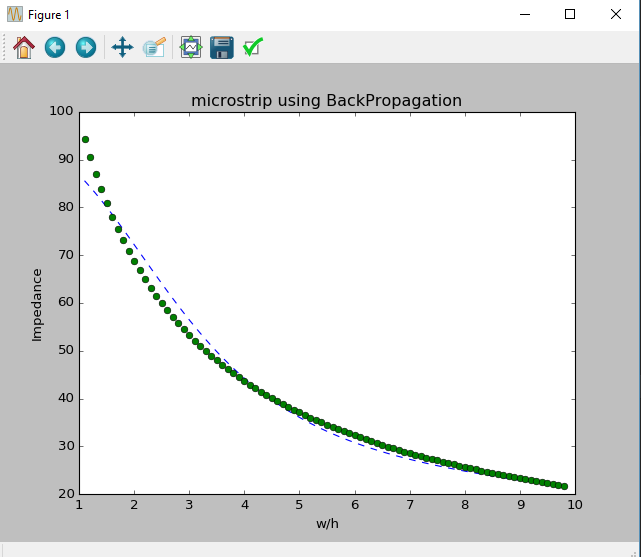}
  \caption{Impedance Vs. w/h for Microstrip Line using neural networks}
  \label{fig:fig1}
\end{figure}

\begin{figure}[H]
  \centering
  \includegraphics[scale=0.4]{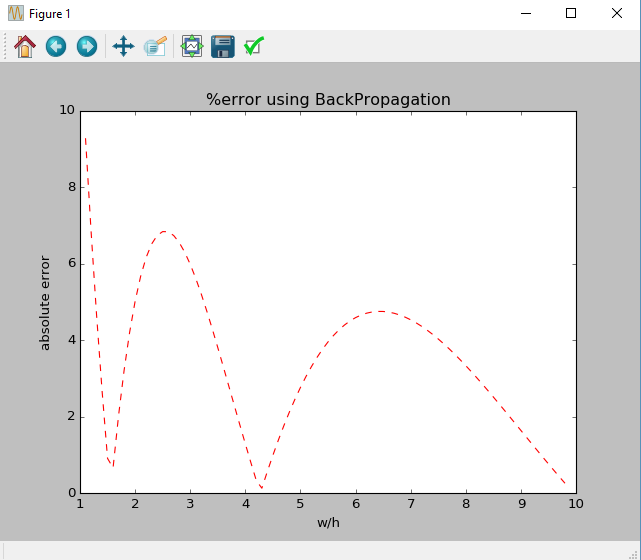}
  \caption{ Absolute Error Vs. w/h for Microstrip Line using  neural networks}
  \label{fig:fig2}
\end{figure}

\begin{figure}[H]
  \centering
  \includegraphics[scale=0.4]{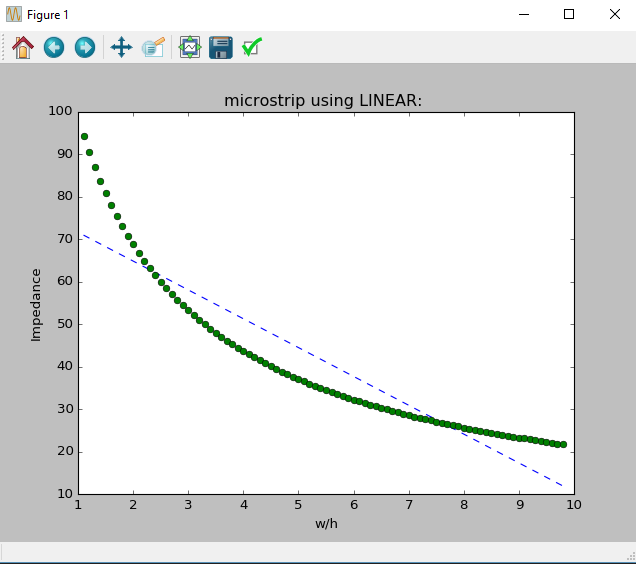}
  \caption{Impedance Vs. w/h for Microstrip Line using linear regression}
  \label{fig:fig3}
\end{figure}

\begin{figure}[H]
  \centering
  \includegraphics[scale=0.4]{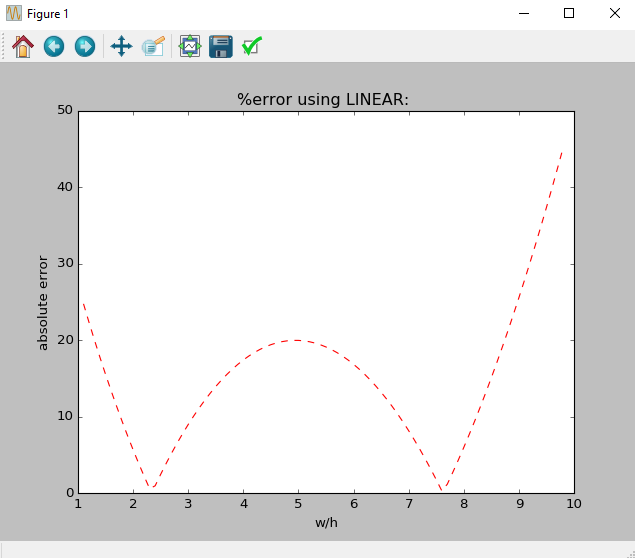}
  \caption{ Absolute Error Vs. w/h for Microstrip Line using  linear regression}
  \label{fig:fig4}
\end{figure}

\begin{table}[ht]
\centering
\caption{Predicted Impedance values for Microstrip Line using neural networks and Linear Regression}
\label{tab:impedance_error}
\begin{tabular}{rrrrrr}
\toprule
w/h & Impedance (actual) & neural networks & Linear Regression & \% Error (neural networks) & \% Error (Linear Regression) \\
\midrule
1.000 & 98.525 & 97.323 & 71.658 & 1.220 & 27.269 \\
1.500 & 80.819 & 81.450 & 68.265 & 0.781 & 15.533 \\
2.000 & 68.774 & 68.488 & 64.871 & 0.416 & 5.675 \\
2.500 & 59.999 & 60.280 & 61.477 & 0.468 & 2.463 \\
3.000 & 53.296 & 53.129 & 58.084 & 0.313 & 8.984 \\
3.500 & 47.993 & 48.268 & 54.690 & 0.573 & 13.954 \\
4.000 & 43.686 & 43.408 & 51.297 & 0.636 & 17.422 \\
4.500 & 40.113 & 39.950 & 47.903 & 0.406 & 19.420 \\
5.000 & 37.098 & 37.376 & 44.509 & 0.749 & 19.977 \\
5.500 & 34.517 & 34.802 & 41.116 & 0.826 & 19.118 \\
6.000 & 32.281 & 32.228 & 37.722 & 0.164 & 16.855 \\
7.000 & 30.325 & 29.801 & 34.329 & 1.728 & 13.204 \\
7.500 & 28.598 & 28.517 & 30.935 & 0.283 & 8.172 \\
8.000 & 27.061 & 27.233 & 27.541 & 0.636 & 1.774 \\
8.500 & 25.685 & 25.949 & 24.148 & 1.028 & 5.984 \\
9.000 & 24.445 & 24.665 & 20.754 & 0.900 & 15.099 \\
\bottomrule
\end{tabular}
\end{table}

\subsection{Microstrip Patch Antenna}
In this study, the rectangular microstrip antennas consist of a rectangular patch of varying dimensions, namely width (W) and length (L), positioned over a ground plane separated by a substrate with thickness (h). The substrate material's dielectric constant \(\varepsilon_r\) plays a critical role in determining the antenna's properties. For the purpose of the analyses presented, a dielectric constant \(\varepsilon_r\) of 6 is used, which falls within the typical range for microstrip antenna substrates \(2.2 <\varepsilon_r < 12\). All plots and results are derived using this specified dielectric constant.

\subsection{Results for Patch Antenna}
\begin{figure}[H]
  \centering
  \includegraphics[scale=0.4]{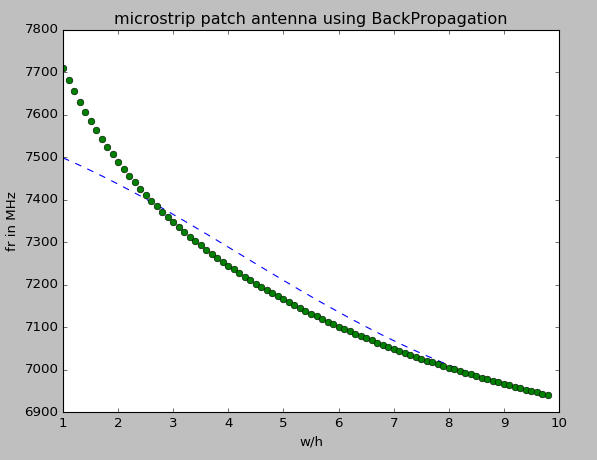}
  \caption{Impedance Vs. w/h for Patch using neural networks}
  \label{fig:fig5}
\end{figure}

\begin{figure}[H]
  \centering
  \includegraphics[scale=0.4]{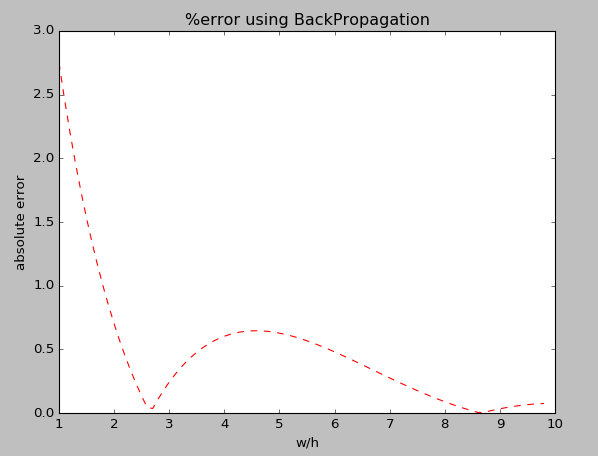}
  \caption{ Absolute Error Vs. w/h for Patch using  neural networks}
  \label{fig:fig6}
\end{figure}

\begin{figure}[H]
  \centering
  \includegraphics[scale=0.4]{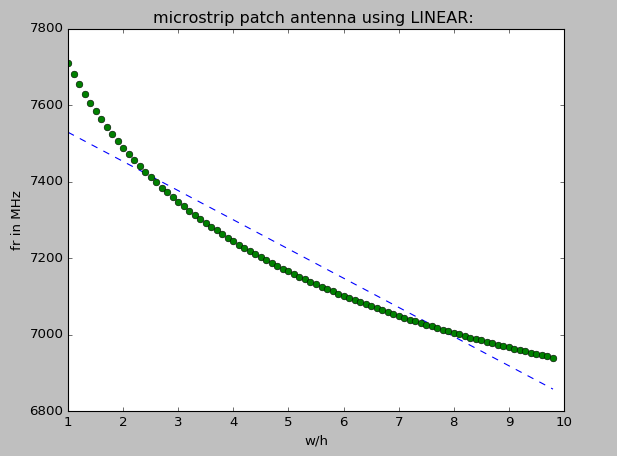}
  \caption{Impedance Vs. w/h for Patch using linear regression}
  \label{fig:fig7}
\end{figure}

\begin{figure}[H]
  \centering
  \includegraphics[scale=0.4]{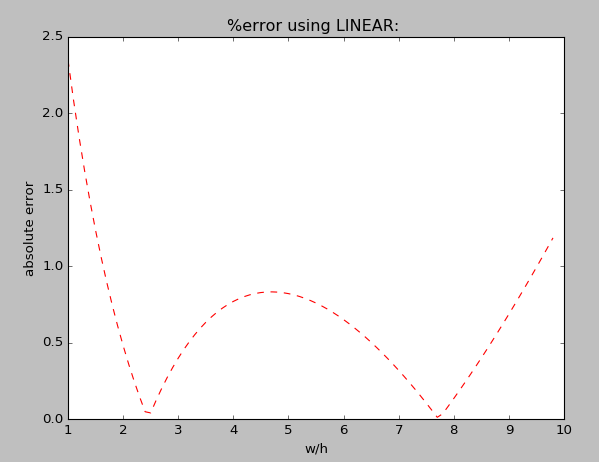}
  \caption{ Absolute Error Vs. w/h for Patch using  linear regression}
  \label{fig:fig8}
\end{figure}

\begin{table}[ht]
\centering
\caption{Frequency vs. w/h for Microstrip Line with Error Percentage}
\label{tab:frequency_error}
\begin{tabular}{rrrrrr}
\toprule
w/h & Fr in MHz (actual) & neural networks & Linear Regression & \% Error (neural networks) & \% Error (Linear Regression) \\
\midrule
1.0 & 7710.557 & 7692.914 & 7529.473 & 0.229 & 2.349 \\
1.5 & 7585.017 & 7593.029 & 7491.309 & 0.106 & 1.235 \\
2.0 & 7489.211 & 7493.143 & 7453.145 & 0.053 & 0.482 \\
2.5 & 7411.944 & 7397.386 & 7414.982 & 0.196 & 0.041 \\
3.0 & 7347.491 & 7348.007 & 7376.818 & 0.007 & 0.399 \\
3.5 & 7292.474 & 7298.629 & 7338.654 & 0.084 & 0.633 \\
4.0 & 7244.707 & 7251.310 & 7300.490 & 0.091 & 0.770 \\
4.5 & 7202.688 & 7204.505 & 7262.326 & 0.025 & 0.828 \\
5.0 & 7165.336 & 7157.707 & 7224.162 & 0.106 & 0.821 \\
5.5 & 7131.843 & 7123.447 & 7185.999 & 0.118 & 0.759 \\
6.0 & 7101.594 & 7100.492 & 7147.835 & 0.016 & 0.651 \\
6.5 & 7074.102 & 7077.537 & 7109.671 & 0.049 & 0.503 \\
7.0 & 7048.982 & 7054.582 & 7071.507 & 0.079 & 0.320 \\
7.5 & 7025.921 & 7031.627 & 7033.343 & 0.081 & 0.106 \\
8.0 & 7004.661 & 7008.672 & 6995.180 & 0.057 & 0.135 \\
8.5 & 6984.987 & 6985.717 & 6957.016 & 0.010 & 0.400 \\
9.0 & 6966.719 & 6962.762 & 6918.852 & 0.057 & 0.687 \\
9.5 & 6949.706 & 6939.807 & 6880.688 & 0.142 & 0.993 \\
\bottomrule
\end{tabular}
\end{table}

\section{Analysis}
This study has systematically investigated various neural modeling techniques for predicting the characteristics of transmission lines, specifically examining their applicability and accuracy compared to traditional computational methods. Our analysis utilized actual impedance and frequency measurements from microstrip lines to evaluate the performance of neural networks and linear regression algorithms. The results demonstrated that while no single method uniformly excels across all scenarios, neural networks generally provided closer approximations to actual measurements than linear regression. For instance, in impedance predictions, neural networks showed a maximum error of approximately 0.229\% for a w/h ratio of 1.000, whereas linear regression exhibited a significantly higher error of 2.349\% under the same conditions. Similarly, in frequency estimations, neural networks consistently maintained lower error margins, indicating its superior ability to adapt to the nonlinear dynamics of microstrip line characteristics. The traditional approaches, often implemented through MATLAB toolboxes or similar simulation software, while effective, are somewhat rigid, limiting their adaptability to different antenna architectures. The models discussed in this paper offer a more flexible and scalable alternative. By employing neural networks, we have described synthesis and analysis methods that not only handle varying transmission line parameters but also demonstrate scalability and adaptability to diverse antenna designs.

\section{Conclusion}
Our findings underscore the importance of choosing the right model based on the specific characteristics, requirements and architecture of the transmission line. The errors observed in our simulations were within acceptable limits, affirming the viability of using advanced machine learning techniques for the design and analysis of transmission lines. These insights pave the way for future research to further refine these models, potentially integrating more complex\cite{nguyen2000}\cite{gevorgian2003} transmission lines and diverse datasets to enhance predictive accuracy and reliability across a broader range of transmission line applications.


\begin{thebibliography}{99}

\bibitem{gupta1979}
K.C. Gupta and Ramesh Garg, \emph{Microstrip Lines and Slotlines}, Artech Inc., pp. 43-50, Jun. 1979.

\bibitem{nguyen2000}
Cam Nguyen, \emph{Analysis Methods for RF, Microwave, and Millimeter-Wave Planar Transmission Line Structures}, Texas A and M University, John Wiley and sons, Inc., pp. 68-80, Aug. 2000.


\bibitem{singh2015}
Bablu Kumar Singh, ``Design of Rectangular Microstrip Patch Antenna based on Artificial Neural Network Algorithm,'' Jodhpur Institute of Engineering and Technology, 2nd International Conference on SPIN, 2015.

\bibitem{gevorgian2003}
S. Gevorgian, H. Berg, H. Jacobsson, and T. Lewin, ``Application Notes Of Basic Parameters Of Coplanar-Strip Waveguides On Multilayer Dielectric/Semiconductor Substrates,'' \emph{IEEE Microwave Magazine}, vol. 4, issue 2, Jun. 2003.

\bibitem{naser-moghaddasi2007}
M. Naser-Moghaddasi, Pouya Derakhshan Barjoei, Alireza Naghsh, ``A Heuristic Artificial Neural Network for Analyzing and synthesizing Rectangular Microstrip Antenna,'' \emph{IJCSNS International Journal of Computer Science and Network Security}, vol.7, no.12, Dec. 2007.

\end{thebibliography}
\end{document}